\documentclass{PoS}

\usepackage[authoryear,square]{natbib}
\bibpunct{(}{)}{;}{a}{}{,}

\title{Incoherent transient radio emission from stellar-mass compact objects in the SKA era}

\ShortTitle{Incoherent transients}

\author{\speaker{S. Corbel}$^{1}$\thanks{Co-chair of the chapter.}, J. C. A. Miller-Jones$^{2}$\thanks{Co-chair of the chapter.}, R. P. Fender$^{3}$,  E. Gallo$^{4}$, T. J. Maccarone$^{5}$, T. J. O'Brien$^{6}$, Z. Paragi$^{7}$, M. P. Rupen$^{8}$, A. P. Rushton$^{3}$, S. Sabatini$^{9}$, G. R. Sivakoff$^{10}$, J. Strader$^{11}$ and P. A. Woudt$^{12}$\\
\llap{$^1$}University Paris Diderot \& CEA Saclay, 
\llap\ {$^2$}ICRAR, Curtin University, 
\llap\ {$^3$}Oxford University, 
\llap\ {$^4$}University of Michigan, 
\llap\ {$^5$}Texas Tech University,
\llap\ {$^6$}University of Manchester, 
\llap\ {$^7$}JIVE, 
\llap\ {$^8$}NRAO,
\llap\ {$^9$}INAF,
\llap\ {$^{10}$}University of Alberta,
\llap\ {$^{11}$}Michigan State University,
\llap\ {$^{12}$}University of Cape Town\\
E-mail: \email{stephane.corbel@cea.fr}, \email{james.miller-jones@curtin.edu.au}}

\abstract{The universal link between the processes of accretion and ejection leads to the formation of jets and outflows around accreting compact objects. Incoherent synchrotron emission from these outflows can be observed from a wide range of accreting binaries, including black holes, neutron stars, and white dwarfs. Monitoring the evolution of the radio emission during their sporadic outbursts provides important insights into the launching of jets, and, when coupled with the behaviour of the source at shorter wavelengths, probes the underlying connection with the accretion process.  Radio observations can also probe the impact of jets/outflows (including other explosive events such as magnetar giant flares) on the ambient medium, quantifying their kinetic feedback.

The high sensitivity of the SKA will open up new parameter space, 
enabling the monitoring of accreting stellar-mass compact objects from their bright, Eddington-limited outburst states down to the lowest-luminosity quiescent levels, whose intrinsic faintness has to date precluded detailed studies. A  census of quiescently accreting black holes will also constrain binary evolution processes.  By enabling us to extend our existing investigations of black hole jets to the fainter jets from neutron star and white dwarf systems, the SKA will permit comparative studies to determine the role of the compact object in jet formation.  The high sensitivity, wide field of view and multi-beaming capability of the SKA will enable the detection and monitoring of all bright flaring transients in the observable local Universe (within $\sim$ 15 Mpc), including the radio counterparts of ultraluminous X-ray sources, improving our understanding of accretion and jet ejection at the highest rates, with important implications for the growth of the first quasars. 

As synchrotron events peak earlier at higher frequencies, and with higher flux densities, such studies will be best enabled by SKA1-MID, in the higher-frequency bands 4 and 5.  
With the high sensitivity available from SKA1-MID, we will also be able to probe isolated quiescent black holes undergoing Bondi-Hoyle accretion from the nearby environment, both stellar-mass black holes in the field and the putative population of intermediate black holes in globular clusters.  
This chapter  reviews the science goals outlined above, demonstrating the progress that will be made by the SKA in studying incoherent synchrotron emission from accreting compact objects. We also discuss the potential of the
astrometric and imaging observations that would be possible should a significant VLBI component be included in phase 1 (and eventually phase 2) of the SKA.
}

\FullConference{
Advancing Astrophysics with the Square Kilometre Array\\
June 8-13, 2014\\
Giardini Naxos, Italy}

\def\gx{GX~339$-$4}

\def\1e{1E~1740.7$-$2942}

\def\vq{V404~Cyg}

\def\a0{A~0620$-$00}

\def\1h{H~1743$-$322}

\def\4u{4U~1755$-$33}

\def\sw{Swift J1753.5$-$0127}

\newcommand{\mnras}{MNRAS}
\newcommand{\apjl}{ApJ}
\newcommand{\apj}{ApJ}

\newcommand{\aj}{AJ}
\newcommand{\nat}{Nature}
\newcommand{\aap}{A\&A}

\newcommand{\nar}{New Astron. Rv.}
\newcommand{\pasa}{PASA}  
\newcommand{\aapr}{AA Rv.}  

\newcommand{\etal}{et al \,}

\begin{document}

\section{Introduction and context}
Stellar-mass compact objects provide important laboratories for studying the fundamental coupling between the accretion process and the launching of energetic outflows, which often take the form of highly-collimated jets \citep{Hughes91}. A theoretical picture has been developed where the jets are composed of an electron/positron or electron/proton  plasma \citep[e.g.][]{Bonometto71}, which is magnetically collimated as it flows away from the compact object.  These jets may be powered either by tapping the energy of a rotating black hole \citep{Blandford77}, or by extracting  energy from the accretion flow \citep{Blandford82}. However, many basic aspects of jet physics are uncertain, including their composition and structure, as well as the mechanisms that power and collimate them. Relativistic jets are relevant in almost all fields of astrophysics, and in some cases may be the dominant output channel for the accretion power from black holes \citep{Fender03}. They provide an important source of feedback to the surrounding environment, being able to either trigger or suppress star formation, accelerate cosmic rays, and seed the surrounding medium with magnetic fields. They have even been suggested to play a role in the reionisation of the Universe \citep[e.g.][]{Mirabel11}.

Despite their relative proximity, the lower masses of stellar-mass compact objects imply that they are observed at lower angular resolution (in terms of gravitational radii) than nearby AGN, yet they evolve through their duty cycles on human timescales, typically undergoing entire outbursts over periods of days to months.  Thus, they provide unique insights into the coupling between accretion and outflow.  Furthermore, comparative studies of the different classes of compact object can provide important insights into the necessary and sufficient ingredients for the jet launching processes.

The non-thermal radio emission from stellar-mass compact objects typically arises via synchrotron emission from relativistic particles spiraling around the magnetic field lines of the jets.  While significant progress has been made over the past few decades in understanding the nature of the jets and their coupling to the accretion flow \citep[see][for a review]{Fender06b}, investigations have been hampered by the limited sensitivity of past and current facilities (see also \citealt{Fender14}).  We detail in this chapter how the sensitivity improvement brought about by the SKA will allow us to study black hole outbursts throughout the Galaxy and out into the Local Group, determine the role of jets in the low-luminosity quiescent state, and extend our existing studies of black holes to the analogous yet fainter neutron star and white dwarf systems.
 
 \section{Galactic black holes}  
 \label{sect_gbh}
  
From an observational standpoint, black hole transients spend most of their time in a quiescent state, at very low mass accretion rates.  They occasionally undergo outbursts that last from a few months to $\sim$ a year, during which the flux rises by several orders of magnitude across the whole electromagnetic spectrum \citep{McClintock06}. These outbursts are associated with global changes in three main components: the jets, the accretion disk and the corona. The luminous outburst phase, with a  luminosity $>$ 10--30\% of the Eddington luminosity (the ``soft state''), is dominated by thermal emission from the accretion disk. During the rise and decay phases of the outburst (the ``hard'' state), the bolometric luminosity of the source is dominated by non-thermal emission (synchrotron or inverse Compton emission from either the jets or the corona) extending up to the hard X-ray band. Following  recent results with high-resolution X-ray spectroscopy, accretion disk 
winds are now recognized as ubiquitous in black hole X-ray binaries,  and may carry away a significant fraction of the inflowing energy \citep{MillerJ06d,Neilsen09, Ponti12, DiazTrigo13}. 

Observations of typical Galactic black holes have indicated two forms of jets associated with these primary accretion states; the slowly-varying, partially self-absorbed compact jets (with radio emission usually $\leq$ 30-50 mJy for distances of a few kpc, and a flat or slightly inverted radio spectrum) observed in the hard state \citep{Corbel00, Fender00, Dhawan00,Stirling01}, and the bright (0.1--10 Jy, with an optically thin radio spectrum), strongly variable transient jets (occasionally showing apparent superluminal motion) detected during the transition from the hard to the soft state \citep{Corbel04b,Fender04b}. The core radio emission is then strongly quenched during the soft state \citep{Fender99a, Coriat11}. Relic radio emission can also be detected in some cases when the jets interact with the ambient medium, either as large-scale lobes \citep{Gallo05a}, or as faint, transient hot spots \citep{Corbel02b}, depending on the duty cycle of the central black hole. 

\begin{figure}[t]
\vspace{-0.4cm}
\centerline{\includegraphics[width=.85\textwidth]{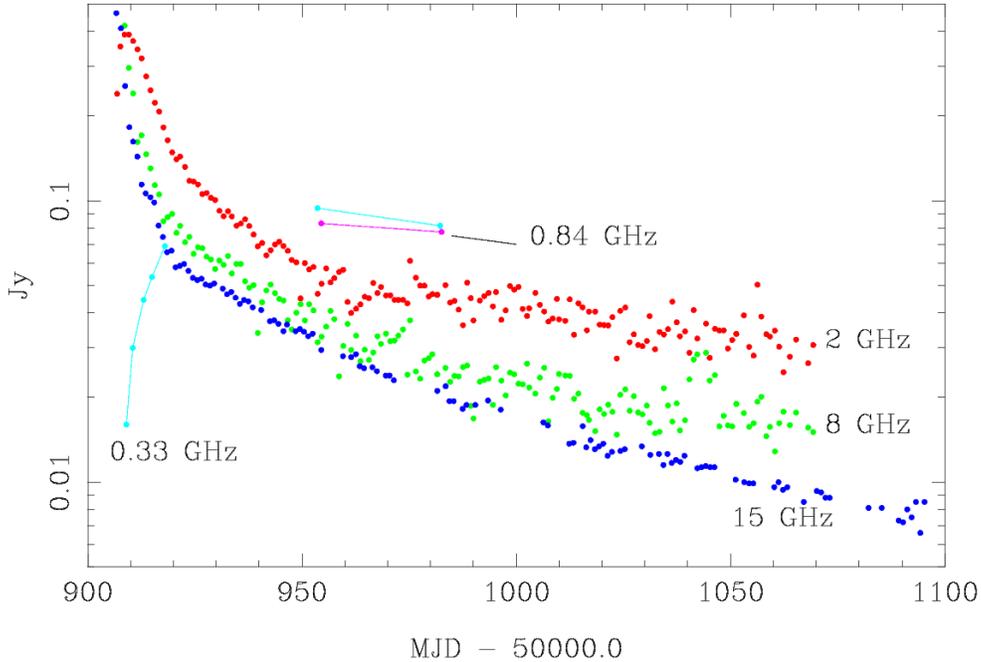}}
\vspace{-0.3cm}
\caption{Radio lightcurve of CI Cam  (a B[e]  X-ray transient, also called XTE~J0421$+$560) following its discovery. As for all incoherent synchrotron transients, the lower-frequency radio emission peaks later and with lower amplitude (Figure from \citealt{Fender08}). Depending on the rate at which a source evolves through its X-ray spectral states, such radio flares typically occur a few days to months following the detection of the X-ray outburst.}
\label{fig_cicam}
\end{figure}
As is characteristic of expanding synchrotron sources, the bright radio flares from the transient jets display a characteristic frequency dependence of the transition (see Fig.~\ref{fig_cicam}) from optically-thick to optically-thin emission, which is  consistent with an expanding plasmon model such as the one described by \citet{VdL66}. 
Thus, the low-frequency radio emission ($<1$\,GHz) will peak later (usually by a few days to weeks) and at lower fluxes (by at least an order of magnitude) compared to the emission at higher frequencies ($>3$ GHz). To study the transient radio emission, it is therefore of the utmost importance for the initial deployment of SKA1-MID to include band 4  (2.80-5.18 GHz) or (preferably) band 5 (4.6-13.8 GHz). This is also important to constrain the energy budget of the compact jets seen in the hard state, which can be strongly affected by opacity effects at lower frequencies.  Bands 4 and 5 have higher intrinsic sensitivity than bands 1--3 (see Table 7 of the SKA1 System Baseline Design Document; band 1 = 0.35-1.05 GHz, band 2 =0.95-1.76 GHz, band 3 = 1.65-3.05 GHz),  as well as lower confusion limits and higher intrinsic levels of radio emission at the peak of an outburst (see Fig.~1 or any textbook example of a synchrotron flare; e.g.\  \citealt{VdL66}).  This allows them to probe a larger observing volume than the lower frequency bands, by factors of between 6 (comparing band 4 to band 3) and 200 (comparing band 4 to band 1), thereby extending the maximum distance we can probe by factors of 2--6.  The brighter peak emission in band 5 would increase these numbers still further.

\begin{figure}[t]
\vspace{-0.4cm}
\centerline{\includegraphics[width=.85\textwidth]{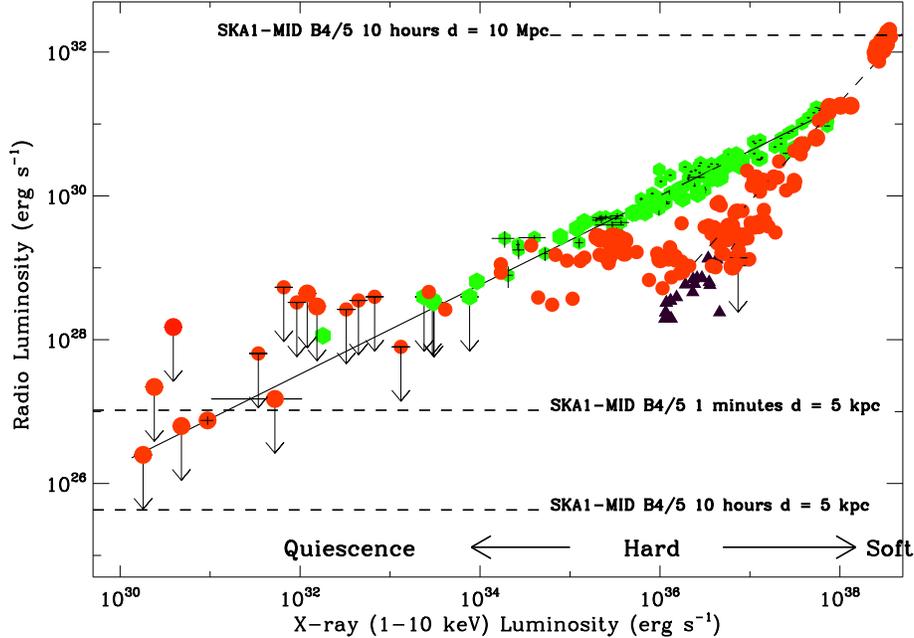}}
\vspace{-0.2cm}
\caption{Radio  and X-ray (1-10 keV) luminosities for Galactic accreting binary black holes in the hard and quiescent states (see arrows at bottom for approximate luminosity levels). 
It illustrates  the standard correlation (defined by sources such as \gx\ or \vq\ with index $\sim$ 0.6; green points) and the new correlation for the so-called ``outliers'' (defined by e.g.  \1h\ or \sw\ with index $\sim$ 1.4; a significant fraction of the red points). Black triangles represent neutron star binaries.  The solid line illustrates the  fit to the whole 1997--2012 sample of \gx\ (\citealt{Corbel13})
with an extrapolation to the quiescent state. The dotted line corresponds to the fit to the data for \1h, one of the representatives 
for the outliers \citep{Coriat11}. Upper limits are plotted at the 3$\sigma$ confidence level. The horizontal dashed lines represent some sensitivity 
levels for SKA1-MID as discussed in the text. }
\label{fig_correl}
\end{figure}

There is a strong correlation between the radio and X-ray emission in the hard and quiescent states \citep{Corbel03, Corbel13, Gallo03, Gallo12}, 
indicating a possible coupling between  the launch of the jets and the dynamics of the flow close to the accreting black hole (Fig. \ref{fig_correl}). 
It can be used to determine the expected source behaviour at the lowest radio luminosities, which remain relatively poorly explored owing to the limited sensitivity of current instruments \citep{MillerJones11}. It is, for example, not clear whether quiescent black holes do host jets as in the brighter states. Current observations seem to indicate weak jet activity even in quiescence \citep{Gallo06,Gallo14}, but it is not known what fraction of the liberated accretion power is carried away in the jets rather than being advected across the black hole event horizon \citep{Fender03}.   Furthermore, understanding the recently-discovered dichotomy in the radio/X-ray correlation \citep{Coriat11,Gallo12} could provide a new way to explore the possibility of different couplings between accretion and ejection in black hole transients.   With a sensitivity as good as 30 $\mu$Jy s$^{-1/2}$ in band 4/5 \citep{Dewdney13}, SKA1-MID is the ideal instrument to monitor all detected black hole outbursts (up to a few tens per year) all the way down to quiescence. Furthermore, a long observation (10 hrs) with SKA1-MID could detect all quiescent black hole binaries up to a distance of 5\,kpc. These transient black holes could either be discovered by targetted monitoring or a blind radio survey with SKA \citep{Fender14}, or by any active multi-wavelength monitoring facility.

The brightest phases of black hole outbursts (the hard to soft state transitions) can easily be studied via snapshot observations, even for the most distant systems in our Galaxy, meaning that SKA (even in phase 1) will probe a significant fraction of the outburst activity for almost all black holes in our Galaxy, including the very faint, sub-Eddington transients, which have rarely been observed at radio frequencies. With a resolution on the order of a few tenths of an arcsecond -- if SKA1-MID is equipped with band 4 or 5 receivers -- the bright radio emission 
associated with the transient jets \citep{Mirabel94} could also be resolved from a week after the onset of the radio flare, as typical proper motions are on the order of 15--20 mas\,day$^{-1}$ (see Section \ref{sec:vlbi} for the additional benefits of VLBI). Good polarization measurements (linear and circular) from SKA1-MID, associated  with good signal purity, will provide key probes of the composition and geometry of the jets, and the structure of their magnetic fields.

The high sensitivity and ability to dump the visibilities on rapid ($\lesssim1$\,s) timescales would allow us to probe any short-timescale variability of the jets, should it be present (shortest timescales of $\sim$ 10 minutes up to now, e.g. \citet{Corbel00, Middleton13}).  \citet{Casella10} detected rapid (sub-second) infrared variability from the jets of \gx , which was correlated with the observed X-ray variability.  Observing these variations at multiple frequencies as they propagate downstream in the jets would allow us to directly measure the speeds and sizes of the compact jets, providing the best observational constraints on their Lorentz factors.
 
An X-ray binary by nature includes several emission components (star, compact object, accretion disk, corona, jets, and nearby environment), all emitting over a wide energy range (the jets themselves  can be observed to emit up to the gamma-ray band), meaning that SKA observations would greatly benefit from coordinated and simultaneous observations with multi-wavelength facilities \citep[e.g.][]{Rodriguez08a,Corbel13b}.  Some obvious new facilities for transient black holes in the next decade include (but are not limited to) X-ray observatories (e.g. {\em SVOM, ASTROSAT}), ground-based all-sky optical telescopes such as LSST, or a sensitive new high-energy gamma-ray observatory (CTA).  Furthermore, all SKA1 components have the potential to act as radio ``all-sky monitors'' (if commensal transient search capability is implemented in the early phases; see \citealt{Fender14}). This would be of prime importance, allowing SKA to provide alerts to multi-wavelength facilities.  Such a radio sky monitor could be crucial for detecting new outbursts should no X-ray all-sky monitor be available in the coming decades.

With a reduced sensitivity of 50\%  during the deployment of SKA1-MID, a significant fraction of the above scientific goals could still be achieved as long as band 4 or 5 is included in the early phase. If we consider the improved performance of SKA2, with broadband frequency access up to 24 GHz, then it is likely that almost all transient black holes located up to the distance of the Galactic Centre could be studied in great detail, as outlined above. \\

 \section{Neutron star binaries}
 
 Comparisons between accreting neutron stars and  black holes are ideal for determining which effects seen from accreting black holes are fundamentally related  to the presence of an event horizon, instead of being generic to accretion onto objects with deep gravitational potential wells.  Largely speaking, the phenomenology of accretion onto neutron stars is similar to that onto black holes \citep[e.g.][]{Psaltis06}. Many of the known differences can be explained in a straightforward
manner by the presence of a solid surface for neutron stars and the lack of one for black holes.

Studies of accreting neutron stars will be bolstered by the enhanced sensitivity of the SKA.  Systematic studies of the radio luminosities of accreting neutron stars as a function of their X-ray luminosities are far sparser than those for black holes.  This is in part because the characteristic timescales on which accretion disks change scales  inversely  with the accretor mass, so that ``typical'' outbursts of accreting neutron stars are shorter than those for accreting black holes; and partly because both the peak X-ray luminosities of neutron star transients and the ratio of radio to X-ray flux for neutron stars are both lower than for black holes \citep{Fender01d,Wu10b,Migliari06b}.  
As a result of these factors, most soft X-ray transients with neutron star accretors peak at flux densities of about 1 mJy or less (for typical distances of several kpc).  Until recently it was not feasible to probe flux densities significantly fainter than 100 $\mu$Jy, and even with the upgraded VLA, the sensitivity of short target-of-opportunity (ToO) observations is limited to a few $\mu$Jy.  Thus, it has traditionally been difficult to span more than a factor of 10 in radio luminosity for neutron stars, and even now it is nearly impossible to span much more than a factor of 100.
Furthermore, since the outbursts progress quickly, scheduling a large number of epochs places strong pressure on the ToO scheduling  of existing arrays.  Finally, since many of these sources are
located in the Galactic Bulge, or elsewhere in the Southern Hemisphere, often only short observations are possible. 
Despite these problems, some progress has still been made on understanding the radio emission of neutron star X-ray binaries.  A few things seem clear: the high magnetic field X-ray pulsars are not strong radio emitters \citep{Migliari06b}; the radio and X-ray fluxes are correlated (see Fig. \ref{fig_correl}) for neutron star X-ray binaries that emit at less than about 10\% of the Eddington luminosity \citep{Migliari03,Tudose09}; and the reduction of the radio power of neutron stars in accretion states dominated by thermal emission is less extreme than the same turn-down for black holes \citep{Migliari04}.  

The enhanced sensitivity of the SKA should allow more quantitative statements on these topics.  For example, \citet{Migliari03} find that $L_R \propto L_X^{1.4}$ for the neutron star 4U~1728-34 in hard spectral states, while \cite{Tudose09} looked at Aql X-1 and found $L_R \propto L_X^{0.4}$, albeit while including data from a range of X-ray spectral states.  The finding of \cite{Migliari03} made for nice agreement with a theoretical picture in which the bulk power put into the jet scales with the mass accretion rate, and the radio power of the jet scales with its bulk power to the 1.4, and also bolsters the suggestion that most hard states of black hole X-ray binaries are in radiatively inefficient states where $L_X \propto \dot{m}^2$.  Developing a sample of sources that has both many objects, and which spans a few orders of magnitude in radio luminosity would  bring the data quality into line with that from the black hole transients at the present time  (Fig. \ref{fig_correl}).  It also opens the door to making robust determinations of whether the spin period of a neutron star affects the normalization of its radio/X-ray relation \citep{Migliari11}.

Additional studies of the thermal X-ray states of neutron stars in the radio may also hold important clues to understanding how jets are launched.  In these systems, by analogy with the black holes, the accretion disks themselves are unlikely to supply much power to an accretion flow.  On the other hand, the boundary layers, where the neutron star's accretion disk dissipates its excess rotational energy, should have the large scale height thought to be needed to power jets, and may interact with the magnetic field of the neutron star itself. Studies of a large sample of these soft states rather than merely the two that have been detected already \citep{Migliari04}, may help us to understand jet production in this environment.

A final class of systems that can probe the relationship between the accretion flow and the launching of jets are the recently-discovered class of transitional binary pulsars, which switch between accretion-powered and rotation-powered states on timescales of just weeks \citep[e.g.][]{Archibald09, Papitto13, Bassa14}.  These systems have been shown to emit flat or inverted-spectrum radio emission in their accretion-powered states, which is consistent with partially-self absorbed synchrotron jets \citep{Papitto13, Bassa14}.  Intensive radio monitoring of the transitions between these two states would provide unique insights into how the jets are formed and destroyed, and, together with simultaneous X-ray monitoring, would highlight the connection between the jets and the evolving accretion flow.

A second point, unrelated to the disk-jet connection, which can be addressed with neutron star observations is whether particles can be launched in a jet at speeds significantly greater than the escape speed of the accretor.  To date, one example has been identified, Cir X-1 \citep{Fender04c}.  In this system an apparent speed of the jet of $15c$ has been inferred from the appearance of radio emission far from the neutron star itself relatively soon after X-ray flaring episodes.  Since the escape speed from a neutron star is typically $\sim0.3c$ this would imply that the bulk jet speed must be larger than the escape speed for a neutron star by a large factor; however, what has not yet been seen, due to lack of sensitivity and angular resolution with existing Southern radio facilities, is actual motion of the radio-emitting structures at an apparent speed larger than the speed of light.  As distances and inclination angles for many neutron star X-ray binaries will be directly measurable with the SKA, the combination of two-sided proper motions and detections of the counterjets should allow for a clean test of whether the pattern speeds and the bulk motions of jets are equal.  If e.g. the jets are dominated by Poynting flux close to the neutron star, then it would not be surprising for them to appear to move very close to the speed of light, even if the particles energized by the jet move at much slower speeds.  Such measurements represent a unique opportunity to test a major hypothesis from theoretical work that cannot be tested with black holes, where the jet speeds are expected to be close
to the speed of light.

\section{Cataclysmic variables and related objects}

The occurrence of jets and fast collimated outflows is by no means restricted to accreting black holes or neutron stars. Through sensitive and timely observations in the last decade, numerous accreting white dwarfs (in cataclysmic variables, symbiotic stars and supersoft X-ray sources) have shown strong evidence for jets and jet-like shocked, collimated outflows, observed at radio frequencies and interpreted as synchrotron emission. 
However, the brightness temperatures of the radio emission from cataclysmic variables are small enough that thermal and/or cyclotron emission are still viable possibilities, and given the multiple possible mechanisms, there may be some heterogeneity in emission mechanisms.  Better spectral measurements, and searches for circular polarization, which should be possible with SKA, can help resolve these issues.

Perhaps the most striking (and encouraging) aspect is that these transient radio jets have all occurred in the prototypes of subclasses of accreting white dwarfs -- the symbiotic star Z And \citep{Brocksopp04b}, the recurrent nova\footnote{For more on classical and recurrent novae and the SKA, see also the chapter on thermal emission from novae \citep{OBrien14}.} RS Oph \citep{Rupen08} and the dwarf nova SS Cyg \citep{Kording08}, suggesting they are more common in white dwarf accretors than previously assumed. Persistent jets are also thought to occur in nova-like variables \citep{Kording11}; these are systems in a regime of steady high mass transfer rate ($\sim 10^{-9} - 10^{-8}$ M$_{\odot}$ yr$^{-1}$) where accretion onto the white dwarf occurs via a standard thin $\alpha$-disc (see e.g. \citealt{Potter14}).

In close analogy to transient jets in X-ray binaries, the non-magnetic dwarf nova SS Cyg repeatedly exhibits radio outbursts associated with its disc instability outburst cycle \citep{Kording08}. The outbursts in dwarf novae are described by a thermal-viscous instability in the accretion disc, leading to a brief period (days to weeks) of enhanced mass transfer ($\sim 10^{-8}$ M$_\odot$ yr$^{-1}$) onto the white dwarf. Even though radio luminosities are low compared to X-ray binaries -- peak flux densities of $\sim$ 1 mJy at 1--10 GHz for a system at 114 pc \citep[SS Cyg;][]{MillerJones13} -- these systems provide an important link in understanding how accretion is coupled to the outflow of matter across a range of compact accretors. With the expected sensitivities achieved by SKA1-MID we can extend the sample of dwarf novae observed at radio frequencies out to kiloparsec distances, where optical transient surveys such as CRTS and iPTF (and LSST in the SKA era) are finding thousands of new dwarf novae; see \cite{Drake14} for an overview of dwarf novae in CRTS. The sheer numbers of systems, the accessible time scales of the disc instability cycle in cataclysmic variables (weeks to months), and the reasonably well-understood accretion discs around white dwarfs provide an excellent laboratory for the SKA to study accretion physics with targeted (and target-of-opportunity) observations. 

Whilst thermal emission is the dominant component of radio emission in novae \citep{Seaquist08,OBrien14}, a significant number of novae exhibit non-thermal (synchrotron) emission associated with collimated bipolar and jet-like outflows (e.g., RS Oph: \citealt{OBrien06,Rupen08}; V445 Pup: \citealt{Woudt09}; V959 Mon: \citealt{Chomiuk14b}). As recurrent novae are prime candidates for the progenitors of type Ia supernovae, questions surrounding the nature and energetics of the outflow of material during a nova outburst are at the core of the debate as to whether a white dwarf grows in mass during successive nova cycles. Regular, multi-frequency monitoring of Galactic novae with the VLA (an SKA1-MID pathfinder in terms of sensitivity) and e-MERLIN (a pathfinder in terms of angular resolution) is ongoing and will be complemented by deep observations with MeerKAT ahead of SKA1-MID. At the moment we are likely only picking up the tip of the iceberg in terms of synchrotron-emitting novae, i.e.\ those with significant circumstellar medium (e.g. from red giant secondary winds) to provide strong shock interactions, or those sufficiently nearby. A representative census of Galactic novae observed at the sensitivity of SKA1-MID (including band 4 or 5) is required to determine what fraction of novae show evidence for synchrotron emission and to fully understand the processes that lead to the formation of collimated jets following a thermonuclear runaway on the surface of the white dwarf.

\section{Imaging and astrometry with VLBI}
\label{sec:vlbi}

High-cadence, high-resolution imaging using SKA-VLBI \citep{Paragi14} during rare outburst events can provide a wealth of information about the jets in stellar-mass compact objects, and their coupling to the accretion process.  Daily imaging during and after state transitions can determine the proper motions (and any observed deceleration) of the approaching and receding jet ejecta, allowing us to track them back to zero-separation, and hence determine the exact timing of the ejection event.  By comparison with X-ray spectral and timing signatures, this enables us to establish the causal connection between changes in the accretion flow and the ejection of relativistic jets.  Since structural changes in the accretion flow occur on timescales of hours to days, such information cannot be gleaned from radio light curves alone, as both opacity effects and time delays between ejection and shock formation cause a lag between an ejection event and the associated peak in the integrated radio light curve.

From such observations, we can determine the product of the jet speed and inclination angle, $\beta\cos\theta$, which can then be decoupled if the distance can be well constrained.  Should we detect deceleration, we can constrain both the power of the jets and the density of the surrounding medium.  Spatially resolving the polarisation of the jets can provide information on the magnetic field structure, helping to distinguish shocks from steady flow.  Polarization observations of stellar-mass synchrotron sources on mas scales have been very rare because these objects are usually very faint ($\sim$mJy or below), and sensitive VLBI networks have not been easy to arrange at short notice during outbursts (but see \citealt{Tudose07}). A flexible SKA-VLBI array will therefore play an important role here, even during the initial deployment phase of SKA1-MID \citep[cf.][]{Paragi14}.
To date, intensive VLBI monitoring has only been carried out for a small number of outbursts \citep[e.g.][]{Yang11,MillerJones12b,Paragi13}.  However, the peculiarities of the individual sources, together with scheduling considerations and sensitivity limitations have so far precluded the identification of which changes in the accretion flow give rise to the major jet  events.

The great sensitivity and resolving power of SKA-VLBI will also be invaluable for the particularly interesting lower-mass neutron star and white dwarf systems, which show similar patterns of behaviour \citep{Maitra04,Kording08} but are typically significantly fainter than black holes.  The lower radio luminosities of these two classes of system have meant that they have not been as intensively studied to date.  Comparative studies of accretion-ejection coupling across different classes of compact object can provide important insights into the process of jet formation and collimation, determining how the jet properties depend on the depth of the gravitational potential well, and the presence or absence of a stellar surface and magnetic field.

The high-precision astrometry enabled by SKA-VLBI can also play an important role in compact object studies.  
Compact, partially self-absorbed synchrotron jets show distinct structural properties on (sub-)mas scales \citep[see][and references therein]{Paragi13}. While flat-spectrum compact jets are typically observed in the hard state of black hole binaries (see Section  \ref{sect_gbh}), the first (and only well-established) observation of the expected frequency-dependent position shift in the jet peak-brightness distribution (the \lq\lq core shift'') -- due to the different optical depths at different frequencies -- was in SS433 \citep{Paragi99}. Since the synchrotron optical depth depends on jet parameters (e.g. magnetic field strength and relativistic particle density), the jet peak position will also change  as the source gets brighter or fainter (see  \citealt{Rushton12,Paragi13}). SKA-VLBI will be sufficiently sensitive to reliably measure core shifts for a large number of Galactic black hole binaries, given the astrometric accuracy of a few $\mu$arcsec for sources as faint as $\sim$1~mJy. This will provide information on the structural properties of the jets, but also has the potential to constrain the accretion physics, as has recently been shown for AGN \citep{Zamaninasab14}.

SKA-VLBI astrometry would be important for measuring accurate parallaxes of Galactic transient sources. Distances to X-ray binaries are typically uncertain by a factor of $\sim2$ \citep{Jonker04}, rendering relatively uncertain our measurements of luminosities, jet speeds and even black hole spin.  As faint, persistent, yet unresolved radio sources, X-ray binaries detected in the hard or quiescent states make ideal astrometric targets \citep[e.g.][]{Bradshaw99,Reid11,MillerJones09}.  
The high astrometric accuracy feasible with VLBI currently enables the measurement of a trigonometric parallax distance out to several kpc.  While {\it Gaia} will provide accurate distances for relatively bright quiescent systems that are not highly absorbed, VLBI will remain the only feasible alternative for distance determinations for optically-faint systems in the Galactic Plane, and the only means to accurately measure distances  for transients discovered after the end of the {\it Gaia} mission lifetime.

Even for systems that are either too distant or too faint for parallax measurements, proper motions can be measured over a time baseline as short as the few-month duration of an individual outburst.  When combined with the line-of-sight radial velocity (from optical or infrared observations) and an estimate of the source distance, the full three-dimensional space velocity of the system can be determined \citep[e.g.][]{Mirabel01}.  This can shed light on whether it received a natal kick in a supernova, and hence place observational constraints on the formation of black holes and neutron stars.
Subtracting the measured parallax and proper motion signatures from astrometric observations of a given system can provide additional constraints on the physical dimensions of the binary.  With sufficient astrometric accuracy, we can measure the orbital signature of high-mass binary systems, and the size scale of the unresolved jets, from the scatter of the astrometric residuals perpendicular and parallel to the jet axis (see \citealt{MillerJones14} for more details).

\section{Isolated black holes}

\begin{figure}[t]
\vspace{-0.5cm}
\centerline{\includegraphics[width=.85\textwidth]{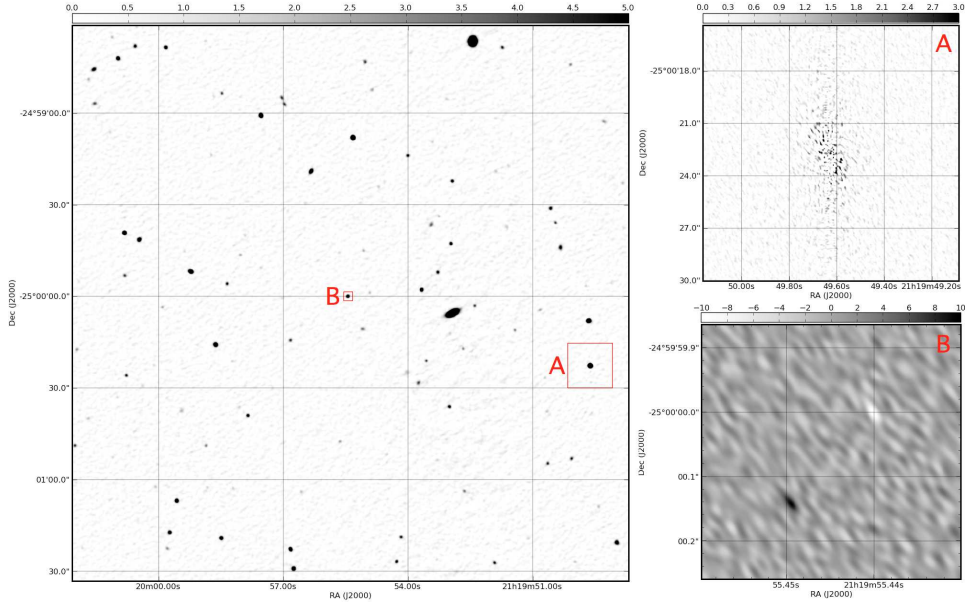}}
\vspace{-0.2cm}
\caption{The main panel shows one of the two simulated SKA2 observations, at 1-2 GHz. Panel A shows the residual emission around a bright source due to incomplete removal of the differing PSFs. Panel B shows a difference image of the region around the IBH, with positive-negative emission due to the motion of the source between the observations. Pixel values are shown with a scale bar above the panel in all cases, the units of which are in $\mu$Jy.beam$^{-1}$. See \cite{Fender13} for more details.}
\label{fig_isolbh}
\end{figure}

There are likely to be $\sim 10^{8}$ stellar-mass black holes in our Galaxy \citep{Samland98}, which, while not in a binary  system, are  still accreting at some low level from the interstellar medium. If the correlation between radio and X-ray luminosities that is observed for stellar-mass black holes in X-ray binaries continues to such low luminosities, then it has been suggested that such sources might be detectable in radio surveys  \citep{Maccarone05}. In \citet{Fender13}, a detailed simulation was performed of a sphere of radius 250 pc centered on the Sun, which is estimated to contain $\sim 35 000$ isolated black holes (IBH), combining our latest understanding of the relation between accretion rate and radio luminosity with the best estimates of the distribution of phases in the ISM. It was found that although the majority of sources would be too faint, a small but significant fraction should be detectable by the SKA as radio point sources with flux densities of a few $\mu$Jy. Such sources would be distinguishable from faint background radio sources (AGN and starburst galaxies) by their relatively high proper motions, which could be up to $100$ mas yr$^{-1}$. Such an approach may be the best method of finding the closest black holes to the Earth. Figure \ref{fig_isolbh}  shows a full simulation of an SKA observation of a field containing such a nearby black hole, demonstrating how it can be identified via its proper motion. Similarly,  a population of floating 
intermediate-mass black holes  may also  be detectable with the SKA, when they traverse the  disk of our Galaxy \citep{Wang14}.

\section{Gamma-ray binaries }

With the advent of very high energy (e.g. MAGIC, HESS) and high energy  (e.g. HESS or Fermi) observatories, a new class of binary systems has been brought to light. These these so-called ``gamma-ray binaries"  are characterized by having most of their radiative power above 1 MeV \citep{Dubus13}. Their non-thermal gamma-ray emission  is driven by rotation-powered pulsars and likely originates from the particles accelerated at the shock front between the wind of a pulsar and the wind of the massive companion.  They constitute ideal testbeds for pulsar physics and the acceleration of particles up to very high energies. 
Variable continuum radio emission has been reported in all of the 5 detected gamma-ray binaries, likely due to  synchrotron emission from electrons accelerated at the shock. Furthermore, in three of these systems, the radio source was resolved on milli-arcsecond scales \citep{Dhawan06,Moldon11,Moldon11b,Moldon12},  
illustrating a cometary tail with a changing morphology along the course of the orbit, a key expectation in the pulsar wind scenario.  However, radio pulsations have only been detected in the  system with the longest orbital period, PSR B1259-63.  The sensitivity of the SKA (even with SKA1-MID 50\%)  will not only allow searches for very faint pulsations in these systems, but also reveal in great detail the properties of the radio flare (i.e.\ the onset of particle acceleration) when the neutron star is close to periastron passage.  Furthermore, direct imaging by the SKA-VLBI  of the evolution of the morphology of the cometary tails will be a complementary tool for understanding the environment of the binary and fundamental issues such as particle acceleration. \\

\section{Magnetars}

Magnetars are isolated neutron stars with very large ($\gtrsim10^{13}$G) magnetic fields, often identified by their association with short bursts of gamma-rays. These systems occasionally undergo giant outbursts, probably powered by reconfiguration of the magnetic field, which can be associated with radio flares and spatially-resolved, mildly-relativistic ejecta. The best example to date is that of the magnetar SGR~1806$-$20, which on Dec 27 2004 underwent probably the largest outburst within our Galaxy, observed by humans, since Kepler's supernova. The outburst was associated with a flare that peaked at over 100 mJy at 1.4 GHz, which originated in relativistic ejecta expanding at $\geq 0.3$c \citep{Gaensler05,Fender06a}. Since this source lies at $\sim 10$kpc, the $\sim 10\mu$Jy sensitivity of the SKA1-SUR and -MID (in hours) could potentially detect such events at Mpc distances, i.e.\ surveys of the local group. Such a survey could also be conducted with SKA1-LOW, as the radio flare from SGR~1806$-$20 was already optically thin and bright at low frequencies ($\sim$ 500 mJy at 250 MHz) very early in the outburst. See \citet{Tauris14} for more details of isolated neutron stars.

\section{Extragalactic binaries and ultraluminous X-ray sources}

As discussed above, studies of Galactic X-ray binaries have significantly improved our understanding of the coupling between the processes of accretion and ejection.  However, as the accretion rate approaches the Eddington limit, the increased radiation pressure changes both the structure of the accretion flow and the nature of the associated outflows (both winds and jets; \citealt{Ohsuga11}).  Owing to the relatively small number of Galactic sources reaching such high luminosities, our understanding of this regime is less well advanced, which is unfortunate given its  important implications for the rapid growth of the most massive quasars in the early Universe.  It has even been demonstrated that Eddington-rate accretion onto low-mass black holes could play a significant role in the reionisation of the Universe \citep{Mirabel11,Fragos13}.

The population of ultraluminous X-ray sources (ULXs) in external galaxies provides potential laboratories for studying accretion at the highest rates.  With X-ray luminosities $\geq 1\times10^{39}$\,erg\,s$^{-1}$, these sources (recently reviewed by \citealt{Feng11}) may be ordinary X-ray binaries accreting at or above the Eddington rate, massive stellar-mass black holes ($M\lesssim 100M_{\odot}$), or, more exotically, intermediate-mass black holes (IMBHs; $10^2<M/M_{\odot}<10^4$).
Radio emission has been detected from several such sources, either as jet-blown bubbles around powerful, persistent sources (e.g.\ \citealt{Kaaret03a,Soria14}), or (in rare cases) as compact, transient emission from powerful jets during outbursts of flaring ULXs \citep{Webb12,Middleton13}.  Nebulae inflated by powerful jets can be used as calorimeters, providing the only method of measuring the unseen mechanical power of the jets, averaged over the source lifetime.  To date, such nebulae have been detected out to distances of $\sim10$\,Mpc, but with typical radio luminosities of $10^{35}$\,erg\,s$^{-1}$, they could be detected out as far as 100\,Mpc with SKA1-MID in just a few hours of time.  With typical sizes of 200\,pc, sub-arcsecond resolution would be required to ascertain whether the sources were extended, implying a need for observations in bands 4 or 5.  With sufficient resolution, individual jet ejecta can be resolved within the nebulae, placing constraints on the source duty cycle \citep[e.g.][]{Cseh14}. Such high -resolution imaging may also be crucial to  properly image the nebulae in order to understand their origin (e.g. jets, disk winds). 

The existence of a correlation among radio luminosity, X-ray luminosity and mass of accreting black holes (the ``Fundamental Plane of Black Hole Activity; \citealt{Merloni03,Falcke04}) implies that measurements of the X-ray and radio luminosities of an accreting ULX can be used to infer its mass, assuming that it is in the equivalent of the hard X-ray spectral state.  Thus, should IMBHs exist, their masses can be measured (to within a factor of a few) via deep radio and X-ray observations.  As an example, the lower limit of $9000M_{\odot}$ on the mass of HLX-1 \citep{Farrell09} would imply a radio luminosity of 1.3\,$\mu$Jy\,beam$^{-1}$ at 5\,GHz at an X-ray luminosity of $0.02L_{\rm Edd}$, which would be detected at a $5\sigma$ level by SKA1-MID in 3\,hours of time (see also \citealt{Wolter14} for more details on ULXs).

Galactic studies of Eddington-rate accretion are typically hindered by the uncertain source distances, large absorbing columns in the Galactic plane, and the small number of sources reaching the Eddington luminosity.  Since the low-mass X-ray binary population traces the total stellar mass, extending such studies to nearby massive galaxies will provide new insights into the nature of jet-disc coupling at Eddington accretion rates.  A recent transient ULX in M31 showed luminous, compact radio emission, with evidence for multiple discrete flaring events and variability on timescales as short as a few minutes, reminiscent of that seen in the Eddington-rate Galactic source GRS 1915+105 \citep{Middleton13}.  The sensitivity of SKA1-MID will allow us to detect such events out to the Virgo cluster.

\section{Black holes in globular clusters}

Globular clusters are extremely efficient at forming X-ray binaries relative to other Galactic environments, due to the additional formation channels of tidal capture or three-body interactions.  However, the several dozen black holes formed from the collapse of massive stars early in the cluster lifetime were initially thought to efficiently eject one another via mutual gravitational interactions \citep{Kulkarni93,Sigurdsson93}, such that clusters were not thought to be rich hunting grounds for black holes.  This picture has recently been revised by new theoretical and observational results, finding that old clusters could still contain many black holes.

The first good evidence for cluster black holes was the detection of highly variable X-ray sources in extragalactic globular clusters, whose luminosities exceeded $10^{39}$\,erg\,s$^{-1}$ \citep{Maccarone07}.  More recently, deep radio observations with the newly-upgraded VLA detected flat-spectrum radio sources in the cores of the Galactic globular clusters M22 and M62 \citep{Strader12a,Chomiuk13}, which were interpreted as quiescent accreting black holes in binary systems.  With two such sources detected in M22 alone, the total black hole population in that cluster was estimated as 5--100 \citep{Strader12a}.

\begin{figure}[h]
\vspace{-0.2cm}
\centerline{\includegraphics[width=.85\textwidth]{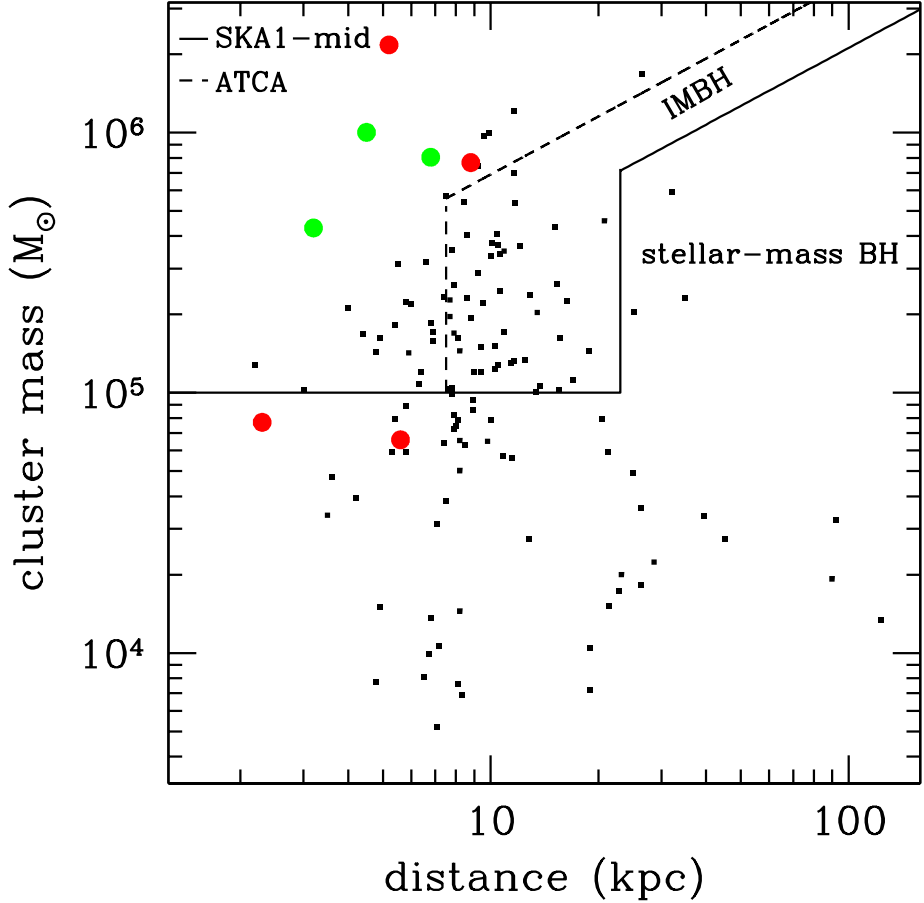}}
\vspace{-0.2cm}
\caption{A summary of all globular clusters in our Galaxy (with declination $<10^\circ$) that could be observed with SKA1-MID. A mass cut of 10$^5$ M$_\odot$ is applied to allow enough black holes to  form early in the cluster lifetime. The green symbols show the detected candidates to date (see references in text), and the red symbols show the observed clusters with no indication of the existence of a  black hole.  The lines illustrate the distance cut corresponding to the distance out to which we can detect the most radio-luminous of our candidate black holes (stellar mass or intermediate mass) at the 5-sigma level for the ATCA (dashed line) or SKA1-MID (full line).  This is $\sim$ 8 kpc for ATCA, and $\sim$ 23kpc for SKA1-MID, for a comparable integration time ($\sim$ 20h observing).  Higher-mass clusters at larger distances are also included, as the best candidates to host IMBHs.}
\label{fig_gc}
\end{figure}

The presence of stellar-mass black holes in globular clusters has also been supported by theoretical work, with $N$-body simulations showing that the interactions between the black hole subsystem in the core and the rest of the cluster would allow a significant number of black holes to persist in the core after a Hubble time \citep{Sippel13,Morscher13,Breen13}.  The additional formation channels for X-ray binaries in clusters implies that these cluster black holes could well be more massive than those in the field, since the progenitor star, if not in a binary system, would have avoided the strong mass loss associated with the common envelope phase of binary evolution.  Thus, in addition to providing new accretion laboratories, these cluster black holes could provide new observational constraints on black hole formation mechanisms.  Finally, the presence of multiple black holes within a cluster core raises the prospect of gravitational wave signals from black hole mergers.

Globular clusters have also been suggested as likely hosts for the elusive IMBHs, which can form in cluster centres from successive mergers of stellar-mass black holes, or via runaway collision of massive stars.  If present, accretion onto these black holes from the intracluster gas should produce detectable radiative signatures \citep{Maccarone05}.  However, while analysis of stellar dynamics has provided some evidence for IMBHs at the centres of several clusters \citep{Gebhardt02,Gerssen02}, there has to date been no good radiative evidence supporting the presence of IMBHs in globular clusters, down to upper limits of several hundred solar masses \citep{MillerJones12c,Strader12b}.
\begin{figure}[b]
\vspace{-0.5cm}
\centerline{\includegraphics[width=.85\textwidth]{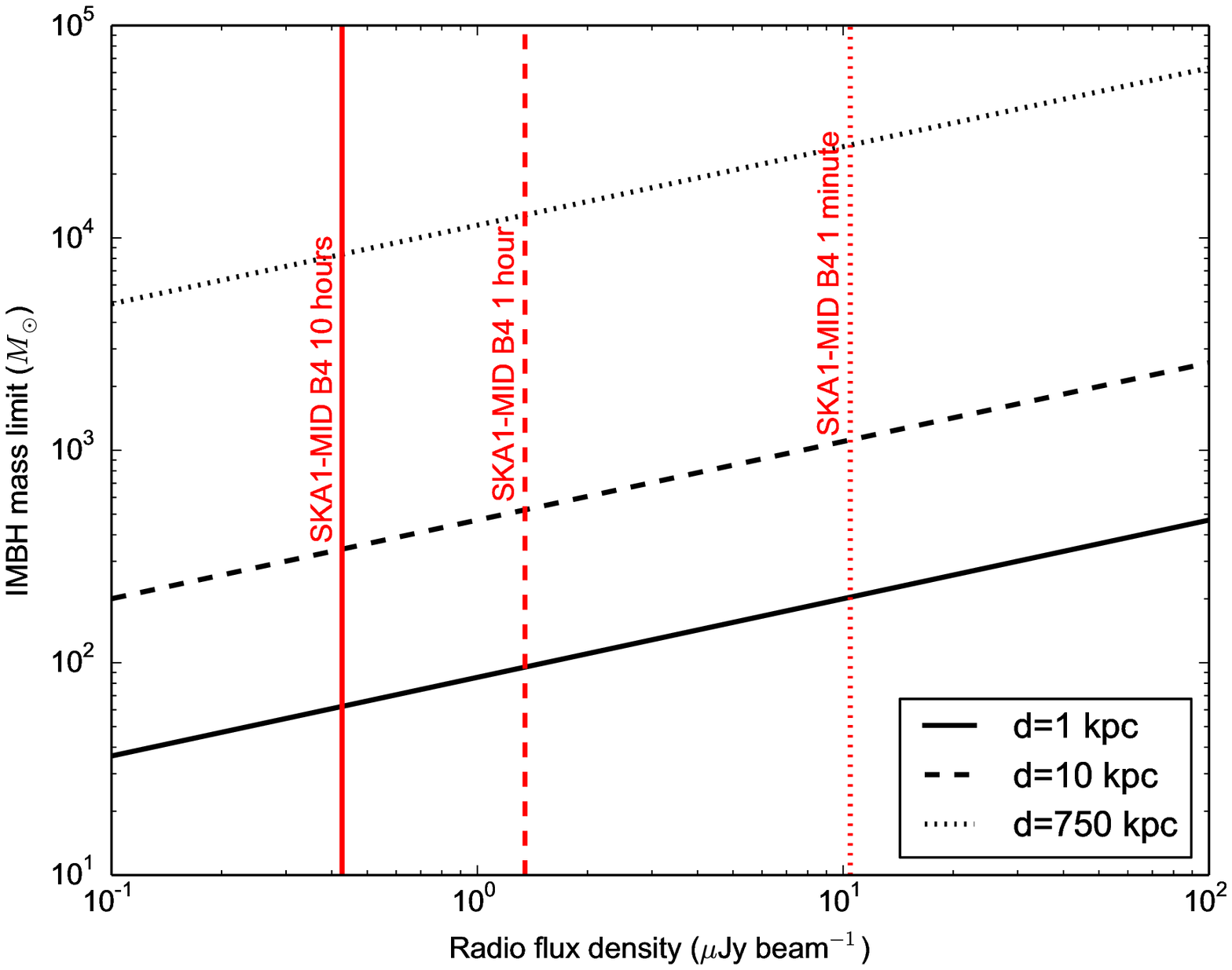}}
\vspace{-0.3cm}
\caption{The SKA can place stringent upper limits on the masses of IMBHs in globular clusters, assuming that the black holes accrete via Bondi-Hoyle accretion from the cluster gas.  Assuming that the accretion leads to the production of jets, we use the formalism of \citet{Strader12b} to calculate the upper limit on the mass as a function of radio luminosity for distances of 1, 10 and 750\,kpc (corresponding to M31).}
\label{fig_imbh}
\end{figure}

With sufficient resolution, the high sensitivity of the SKA can make significant contributions to this field.  The three quiescent black holes detected in M22 and M62 were the first radio-selected stellar-mass black holes.  The well-established, non-linear correlation between radio and X-ray luminosities of accreting stellar-mass black holes in the hard and quiescent X-ray spectral states implies that radio surveys may be more efficient at finding quiescent accreting black holes than X-ray searches \citep{Maccarone05}, as well as providing a way to discriminate between black holes and the less radio-bright neutron star and white dwarf systems.  The most sensitive current radio facility, the VLA, can only reach radio luminosities of $\sim10^{28}$\,erg\,s$^{-1}$ in Galactic clusters, enabling the detection of the brightest quiescent systems (giant donors or ultracompact sources).  A 10-hour run in band 4 using SKA1-MID could drop this limit by an order of magnitude, extending our sensitivity to fainter and more distant quiescent systems (see Fig.~\ref{fig_gc}), and reducing the upper limit on putative IMBH masses to $<150M_{\odot}$ for the most nearby clusters (see Fig.~\ref{fig_imbh}).  Equivalently, the radio outburst of an Eddington-rate X-ray binary flare in an extragalactic cluster could be detected out to the Virgo cluster.  However, sub-arcsecond resolution is required to associate any detected radio sources with the known locations of either X-ray sources (in Galactic clusters) or extragalactic globular clusters.  Sufficiently attenuating confusing steep-spectrum sources such as pulsars or background AGN also strongly motivates conducting such observations in bands 4 or (preferably) 5 on SKA1-MID.

\section{Summary of requirements for the SKA}

As discussed in the previous sections, the most important component of SKA1 for the field of incoherent transients is SKA1-MID, because of its high-frequency capability, implying improved angular resolution, a lower confusion limit, and brighter intrinsic synchrotron emission.
The performance of SKA1-MID (even at 50\% capacity during initial deployment) will largely supersede the current generation of radio interferometers for studies in this field (especially in the Southern hemisphere). However, we stress a number of critical considerations for the final design of SKA1 and its extension to SKA2.
\begin{itemize}
\item Because most of the incoherent transients will initially be optically thick at low frequencies, we emphasise the importance of having band 4 or (preferably)  5 incorporated in the initial deployment of SKA1-MID. This may imply initially dropping band 3, as bands 1 and 2  will be needed to assure continuity with SKA1-LOW and SKA1-SUR, and for the study of the hydrogen in the local Universe. 
\item In addition to having band 4 or 5 included in the initial design, SKA1-MID should be able to operate simultaneously in at least two different frequency bands (possibly via sub-arrays) in order to identify transient radio sources and measure their spectral evolution.  Good polarization capabilities (linear and circular)  are also needed to probe the structure and  composition of the jets. It is important to have 
well-sampled daily radio light curves (with high time resolution) on timescales of  weeks--months, meaning that having sub-arrays or multi-beaming capabilities will greatly enhance the SKA scientific return. 
\item Whereas incoherent transients are not the fastest transients in the sky, it is important to stress that they do show huge variations on timescales of the order of hours (and smaller). It is therefore crucial that
SKA1-MID should be able to respond quickly (few hours/days) in the event of (un)expected transients.  Such a response could be triggered either by SKA itself, or from an external source, as these transients (and also GRBs, SNe, ...) are powerful  
emitters across the entire electromagnetic spectrum.  Relatively short-timescale co-ordination of SKA observations with multi-wavelength facilities should be envisaged and facilitated.   
\item We also stress the importance of commensal searches for radio transients with all SKA components. A dedicated monitoring of the Galactic plane/bulge or selected portions of the sky (perhaps via a radio ``all sky monitor'') 
would facilitate the SKA triggering observations at external observatories, especially if no X-ray all sky monitor is available in the next decade. All new transients should be made public immediately to enable rapid multi-wavelength response. 
\item Angular resolution is crucial for imaging the radio structures associated with those transients, filtering out extended emission to avoid confusion, or studying faint sources in crowded regions of sky such as globular clusters or external galaxies. This can be achieved only if SKA1-MID is equipped 
with band 4 or 5, with the antennas spread over an area of radius a few hundred km. Phasing the SKA dish arrays alone or with existing/planned new VLBI arrays would be a key feature of the SKA, enabling vastly more sensitive imaging and astrometric studies with SKA-VLBI.

\end{itemize}

\bibliographystyle{apj_long_etal}

\end{document}